\documentclass[12pt]{article}

\usepackage{pgfplots}
\usetikzlibrary{decorations.pathmorphing}

\tikzset{snake it/.style={decorate, decoration=snake}}
\pgfplotsset{compat=1.10}
\usepgfplotslibrary{fillbetween}
\usetikzlibrary{patterns}

\usepackage{draft} 
\usepackage{hyperref}
\usepackage{graphicx,color,subfig}
\usepackage{cite}
\usepackage{mciteplus}
\usepackage{skak}
\usepackage{xcolor}
\usepackage{empheq}
\usepackage{tikz}
\DeclareFontFamily{OT1}{pzc}{}
\DeclareFontShape{OT1}{pzc}{m}{it}{<-> s * [1.10] pzcmi7t}{}
\DeclareMathAlphabet{\mathpzc}{OT1}{pzc}{m}{it}

\def\be#1\ee{\begin{align}#1\end{align}}



\newcommand{\gone}[1]{{}}

\newcommand{\N}{{\cal N}}
\newcommand{\M}{{\cal M}}
\newcommand{\V}{{\cal V}}

\newcommand{\beq}{\begin{eqnarray}}
\newcommand{\eeq}{\end{eqnarray}}
\begin{document}

\unitlength = .8mm

\begin{titlepage}

\begin{center}

\hfill \\
\hfill \\
\vskip 1cm

\title{\Huge CFT Correlators from $(0,2)$ Heterotic String}

\author{Amit Giveon$^1$, Akikazu Hashimoto$^2$, David Kutasov$^3$}
\address{
$^1$Racah Institute of Physics, The Hebrew University \\
Jerusalem, 91904, Israel\\
$^2$Department of Physics, University of Wisconsin\\
Madison, WI 53706, USA\\
$^3$Kadanoff Center for Theoretical Physics and Enrico Fermi Institute\\ 
University of Chicago, Chicago IL 60637
}
\vskip 1cm

\email{$^1$giveon@mail.huji.ac.il, $^2$aki@physics.wisc.edu, $^3$dkutasov@uchicago.edu}

\end{center}

\abstract{\noindent In \cite{Giveon:2024fhz}, we argued that the $(0,2)$ heterotic string gives rise in spacetime to left and right-moving symmetric product CFT's. In this paper we confirm this claim by showing that it computes correlation functions in these CFT's. 
}

\vfill

\end{titlepage}

\eject

\begingroup
\hypersetup{linkcolor=black}
\tableofcontents
\endgroup

\vskip 2cm

\section{Introduction}

In this note, we continue our investigation  \cite{Giveon:2024fhz} of the $1+1$ dimensional vacua of the $(0,2)$ heterotic string. In that paper, we showed that the spectrum of this theory on a spatial circle consists of sectors labeled by the winding number $w$. Sectors with positive $w$ contain left-moving excitations. These excitations are described by the symmetric product 
\be\label{mln}
\M_L^{N}/S_{N}~,
\ee
where $\M_L$ is the holomorphic CFT of twenty four left-moving scalar fields whose momenta lie on a Niemeier lattice, and $N$ is a large integer, of order $1/g_s^2$.\footnote{Another posssibility, suggested in a different context in \cite{Aharony:2024fid}, is that the $(0,2)$ string coupling, $g_s$, is related to the chemical potential for $N$, and the relation $g_s^2\sim 1/N$ comes from a saddle point approximation.} The CFT $\M_L$ is isomorphic to the one that enters the worldsheet construction of the $(0,2)$ string. Negative $w$ leads to a right-moving analog of \eqref{mln}. 

As reviewed in \cite{Giveon:2024fhz}, twisted sectors in the CFT \eqref{mln} are labeled by conjugacy classes, $(w_1,\cdots, w_m)$, with $\sum_j w_j=N$. Each $w_j$ can be thought of as corresponding to a $\mathbb{Z}_{w_j}$ twisted sector in $(\M_L)^{w_j}/\mathbb{Z}_{w_j}$. In the string theory, the different conjugacy classes correspond to multi-string sectors consisting of $m$ strings with winding numbers $(w_1,\cdots, w_m)$. The sector with $w=0$ contains the chiral algebra of $\M_L$. 

According to the proposal of \cite{Giveon:2024fhz}, one should be able to calculate correlation functions in the CFT \eqref{mln} using the $(0,2)$ string. At first sight this seems surprising. Indeed, the $\N=2$ string (see e.g. \cite{Marcus:1992wi} for a review) is famous for having vanishing scattering amplitudes, so the mere fact that this particular version of it should have non-vanishing correlation functions appears to be contrary to the lore in the subject. Even if the $(0,2)$ string amplitudes are non-zero, it seems surprising that they would give (holomorphic) two-dimensional CFT correlators, without invoking any weak-strong coupling duality.   

The goal of this note is to show that the above expectations are  realized. In particular, we will show that string amplitudes that correspond to matrix elements between arbitrary incoming and outgoing states with a specific positive winding $w$ match the corresponding amplitudes in the CFT \eqref{mln}. This provides a check on the statement that the dynamics of the left-moving modes in the $(0,2)$ string is described by the left-moving CFT \eqref{mln}. 

Sectors with negative $w$ contain a right-moving analog of the above structure. As mentioned in \cite{Giveon:2024fhz}, when quantum corrections are included, the left and right-movers appear to couple via a $T\bar T$ interaction, with coupling $\alpha'g_s^2$. This aspect of the dynamics will be left for future work. 

In the rest of this note, we will use extensively the results of \cite{Giveon:2024fhz}. Thus, it is best read in conjunction with that paper, which also has further references to relevant literature.
We start in the next section with a review of correlation functions in the CFT \eqref{mln}, and then move on to the string theory calculations.

\section{Correlation functions in $\left(\M_L\right)^N/S_N$}
\label{sec2}

The purpose of this section is to describe (some of) the correlation functions in the CFT \eqref{mln} in a form that will be useful later, when we turn to string theory. We start by discussing the untwisted sector of the orbifold, and then move on to the twisted sectors.

\subsection{$\M_L$}
\label{subsec21}

Consider an operator with  scaling dimension $\Delta\in\mathbb{Z}_+$ in the Niemeier CFT, $V_\Delta(x)$. In general, there are many such operators at a given value of $\Delta$; at large $\Delta$ their number is given by the Cardy formula with $c=24$. We will take $V_\Delta$ to be a Virasoro primary, following \cite{Giveon:2024fhz}, but much of the discussion below is applicable (or easily generalized) to arbitrary $V_\Delta$.

We can pick a preferred basis, in which the two-point function is diagonal, i.e.
\be
\label{twoptf}
\langle V^*_\Delta(y) V_\Delta(x)\rangle\equiv
\langle0| V^*_\Delta(y) V_\Delta(x)|0\rangle={D\over (x-y)^{2\Delta}}~,
\ee
and the two-point function of $V_\Delta$ with any other operator of the same dimension vanishes. The coefficient $D$ in \eqref{twoptf} can be absorbed in the definition of $V_\Delta$, but for comparison with string theory, it will be useful to keep it general. 

We can use the state-operator correspondence to define the state corresponding to the operator $V_\Delta$,
\be
\label{stateV}
|V_\Delta\rangle=\lim_{x\to 0} V_\Delta(x)|0\rangle~,
\ee
where $|0\rangle$ is the $SL(2,\mathbb{R})$ invariant vacuum of the Niemeier CFT. The adjoint of \eqref{stateV} is 
\be
\label{outstateV}
\langle V_\Delta|=\lim_{x\to\infty} x^{2\Delta}\langle 0|V_\Delta^*(x)~. 
\ee
The factor $x^{2\Delta}$ has to do with the conformal transformation $x\to 1/x$ that relates \eqref{stateV} and \eqref{outstateV}. In terms of these definitions, \eqref{twoptf} can be written as 
\be
\label{normV}
D=\langle V_\Delta|V_\Delta\rangle~.
\ee
Unitarity of the Niemeier CFT implies that $D>0$. 

Since all the operators in the Niemeier CFT are holomorphic, one can mode expand them:
\be
\label{modeexp}
V_\Delta(x)=\sum_{n=-\infty}^\infty V_{\Delta,n} x^{-n-\Delta}~,
\ee
whose inverse is
\be
\label{modeexpcont}
V_{\Delta,n}=\oint{dx\over 2\pi i} x^{n+\Delta-1}V_\Delta(x)~.
\ee
As is clear from \eqref{modeexp} and \eqref{modeexpcont}, the modes $V_{\Delta,n}$ have scaling dimension $-n$. Plugging \eqref{modeexp} into \eqref{stateV}, we conclude that it must be that 
\be
\label{killvacright}
V_{\Delta,n}|0\rangle=0 \;\;\;{\rm for}\;\;\; n>-\Delta~,
\ee
so that the limit in \eqref{stateV} exists, and 
\be
\label{vstate}
|V_\Delta\rangle=V_{\Delta,-\Delta}|0\rangle~.
\ee
Thus, the state $|V_\Delta\rangle$ has  $L_0=\Delta$. Taking the adjoint of equation \eqref{killvacright}, we get
\be
\label{killvacleft}
\langle 0|V^*_{\Delta,n}=0 \;\;\;{\rm for}\;\;\; n<\Delta~,\;\;\; {\rm and}\;\;\;\langle V_\Delta|=\langle 0|V^*_{\Delta,\Delta}~.
\ee
An example of the above construction, with $\Delta=2$, is the stress-tensor $T(x)$. The modes $V_{\Delta,n}$, \eqref{modeexpcont}, are in this case the Virasoro generators $L_n$, and the operators that satisfy both \eqref{killvacright} and \eqref{killvacleft} are the global conformal generators $(L_{-1}, L_0, L_1)$.

A Niemeier CFT contains a rank twenty four affine Lie algebra $\widehat G$, and one can classify the operators in it with respect to this algebra. To illustrate the construction, we pick a particular $U(1)$ current $K(x)$ in $\widehat G$, and take the operator $V_\Delta$ to be a primary of charge $q$ under this $U(1)$, 
\be
\label{primuone}
K(x) V_\Delta(y)\sim {q\over x-y} V_\Delta(y)~.
\ee
We will comment on the general, non-abelian, case below (see also \cite{Giveon:2024fhz}). 

Consider the three-point function 
\be\label{corthree}
\langle0| V^*_{\Delta}(x_2) K(x) V_{\Delta}(x_1)|0\rangle={q D\over (x-x_1)(x-x_2)(x_1-x_2)^{2\Delta-1}}~,
\ee
where on the right hand side we used \eqref{twoptf} and \eqref{primuone}. The $x$ dependence of \eqref{corthree} is fixed by conformal symmetry. To understand the content of this equation, we can send $x_1\to 0$, $x_2\to\infty$, as in \eqref{stateV} and \eqref{outstateV}. This gives
\be\label{matthree}
\langle V_\Delta| K(x) | V_{\Delta}\rangle={q D\over x}~.
\ee
Decomposing $K(x)$ into modes, as in \eqref{modeexp} and \eqref{modeexpcont},
\be
\label{modeK}
K(x)=\sum_n K_n x^{-n-1}, \;\;\;K_n=\oint{dx\over 2\pi i} x^nK(x)~,
\ee
we see that only the zero mode $K_0$ contributes to \eqref{matthree}, 
\be\label{zerothree}
\langle V_\Delta| K_n | V_{\Delta}\rangle=q D \delta_{n,0}~.
\ee
This is consistent with the fact that
\be\label{knv}
K_n|V_\Delta\rangle=0\;\;\;{\rm and}\;\;\; \langle V_\Delta|K_{-n}=0\;\;\;{\rm for\;\;\;}n>0~,
\ee
which in turn follows from the fact that 
\be
\label{KV}
[K_n,V_{\Delta,m}]=qV_{\Delta,n+m}~, 
\ee
and from equations \eqref{killvacright} and \eqref{vstate}. Of course, \eqref{vstate} and  \eqref{KV} imply $K_0|V_\Delta\rangle=q|V_\Delta\rangle$.

The above discussion can be generalized to the correlation functions 
\be\label{matl}
\langle V_\Delta| K(x_1) K(x_2)\cdots K(x_l) | V_{\Delta}\rangle~.
\ee
The mode expansion \eqref{modeK} implies that these correlation functions are determined by those of the modes $K_n$. Moreover, using the Kac-Moody algebra, we can place all the raising operators $K_{-n}$ to the right of all the lowering operators $K_n$ (with $n>0$ in both cases). Thus, the correlation functions \eqref{matl} are equivalent to 
\be\label{matlmode}
\langle V_\Delta| K_{m_1} \cdots K_{m_s}K_{-n_1}\cdots K_{-n_r} | V_{\Delta}\rangle~,
\ee
where all the $n_j$ and $m_j$ are positive, and $s+r=l$. Energy conservation implies that 
\be\label{encons}
\sum_{j=1}^s m_j=\sum_{j'=1}^r n_{j'}~.
\ee
To compute \eqref{matlmode}, we use the Kac-Moody algebra satisfied by the $K_n$, and \eqref{knv}. For the $U(1)$ case this boils down to Wick contracting the different $K$'s, and reducing the $l+2$ point function to a two-point function. This means that $s=r$ and $l\in 2\mathbb{Z}$. Furthermore, the sets $\{n_i\}$ and $\{m_j\}$ must be the same.

We note for future reference that the correlation functions \eqref{matlmode} can be obtained from \eqref{matl} by using \eqref{modeK}:
\be\label{ointmode}
\langle V_\Delta| K_{n_1} \cdots K_{n_l} | V_{\Delta}\rangle=
\oint_{\gamma_1}{dx_1\over 2\pi i}\cdots \oint_{\gamma_l}{dx_l\over 2\pi i}x_1^{n_1}\cdots x_l^{n_l}\langle V_\Delta| K(x_1) K(x_2)\cdots K(x_l) | V_{\Delta}\rangle~.
\ee
The integrals over $x_j$ run over concentric contours $\gamma_j$, with $\gamma_j$ containing all $\gamma_i$ with $i>j$. In the next section, we will obtain \eqref{ointmode} from the $(0,2)$ heterotic string. 

Note that: 
\begin{itemize}
\item In the above discussion we took $V_\Delta$ to be a primary of Kac-Moody \eqref{primuone}. This did not lead to any loss of generality, since general states in the CFT can be obtained by acting with KM raising operators \cite{Giveon:2024fhz}, and are thus included in the construction \eqref{matlmode}.
\item We also took the current $K(x)$ to be a $U(1)$ KM current. It is easy to extend the discussion to  general, non-abelian, currents, $K^a(x)$. In that case, in the process of commuting the lowering operators $K^a_m$ in \eqref{matlmode} through the raising operators $K^b_{-n}$, we need to use the full non-abelian KM algebra (equation (3.19) in \cite{Giveon:2024fhz}). Diagrammatically, this leads, in addition to the Wick contractions mentioned above, to contributions from cubic vertices. 
\item We took the in and out-states to correspond to the same KM primary operator $V_\Delta$. If the out-state is different from the in-state, the correlators \eqref{matl} and \eqref{matlmode} vanish.
\end{itemize}
A further generalization of the above discussion that one can consider is to replace the currents $K^a(x)$ in \eqref{matl}, and their non-abelian generalizations, by arbitrary operators in the Niemeier CFT, $V_\Delta$. The above discussion goes through, with the mode expansion \eqref{modeK} replaced by the general ones \eqref{modeexp} and \eqref{modeexpcont}. This leads to the calculation of general $l+2$ point functions of the operators $V_\Delta$, 
\be\label{corl}
\langle0| V_{\Delta_{l+1}}(x_{l+1}) \cdots V_{\Delta_0}(x_0)|0\rangle~.
\ee
Sending $x_0\to 0$ and $x_{l+1}\to\infty$ as in \eqref{stateV} and\eqref{outstateV}, and expanding the remaining operators in modes, using \eqref{modeexp} and \eqref{modeexpcont}, we can relate \eqref{corl} to a generalization of \eqref{matlmode} and \eqref{ointmode},
\beq 
\lefteqn{\langle V^*_{\Delta_{l+1}}| V_{\Delta_l,n_l} \cdots V_{\Delta_1,n_1} | V_{\Delta_0}\rangle } \cr
&=& \oint_{\gamma_1}{dx_1\over 2\pi i}\cdots \oint_{\gamma_l}{dx_l\over 2\pi i}x_1^{n_1+\Delta_1-1}\cdots x_l^{n_l+\Delta_l-1}\langle V^*_{\Delta_{l+1}}| V_{\Delta_l}(x_l) \cdots V_{\Delta_1}(x_1) | V_{\Delta_0}\rangle~. \label{ointmodeV}
\eeq
In order to evaluate the left hand side of \eqref{ointmodeV}, we need to compute the OPE algebra of the $V$'s, and use the mode expansion \eqref{modeexp} and \eqref{modeexpcont} to rewrite it as a commutator algebra for the modes $V_{\Delta,n}$. 

Consider, for example, the three-point function of the operators $V_j(x)=e^{i\vec p_j\cdot\vec y}$, $j=1,2,3$. The three-point function \eqref{corl} is in this case given\footnote{Up to a multiplicative constant, that will not play a role in the discussion below.} by 
\be\label{excorl}
\langle0|e^{-i\vec p_3\cdot\vec y(x_3)}e^{i\vec p_2\cdot\vec y(x_2)} e^{i\vec p_1\cdot\vec y(x_1)}|0\rangle={\delta_{\vec p_1+\vec p_2,\vec p_3}\over (x_2-x_1)^{\Delta_{12}}(x_3-x_1)^{\Delta_{13}}(x_3-x_2)^{\Delta_{23}}}~,
\ee
where $\Delta_{12}=\Delta_1+\Delta_2-\Delta_3$, etc, 
the dimensions $\Delta_j$ are given by $\Delta_j=\frac12|\vec p_j|^2$, and  $\delta$ is a Kronecker $\delta$-function. We can now send $x_1\to 0$ and $x_3\to\infty$, as before, and get  
\be\label{exmatel}
\langle \vec p_3|e^{i\vec p_2\cdot\vec y(x_2)} |\vec p_1\rangle={\delta_{\vec p_1+\vec p_2,\vec p_3}\over x_2^{\Delta_{12}}}~.
\ee
Using the mode expansion \eqref{modeexp} for $e^{i\vec p_2\cdot\vec y(x_2)}$, we find that 
\be\label{modeexm}
\langle \vec p_3|\left(e^{i\vec p_2\cdot\vec y(x_2)}\right)_n |\vec p_1\rangle=\delta_{\vec p_1+\vec p_2,\vec p_3}\delta_{n,\Delta_1-\Delta_3}~,
\ee
i.e.\ only the mode with $n=\Delta_1-\Delta_3$ contributes to \eqref{exmatel}. In the next section, we will reproduce equations \eqref{exmatel} and  \eqref{modeexm} from the string theory perspective.

\subsection{$\left(\M_L\right)^w/\mathbb{Z}_w$ \label{subsec22}}

As reviewed in \cite{Giveon:2024fhz}, the symmetric product CFT $(\M_L)^N/S_N$ has a large set of twisted sectors, labeled by the conjugacy classes of $S_N$. A subset of these sectors corresponds to the $\mathbb{Z}_w$ twisted sectors in $(\M_L)^w/\mathbb{Z}_w$ for various values of $w\le N$. In this subsection, we will discuss this subset. In the next section, we will compare the results of this discussion to $\N=2$ heterotic string theory. 

The untwisted sector operators in $(\M_L)^w/\mathbb{Z}_w$ can be written as 
\be 
\label{untwistop}
\V_\Delta(x) = {1 \over w} \sum_{j=1}^w V_\Delta^{(j)}(x)~, \ee
where $V_\Delta(x)$ is an operator in $\M_L$. The index $j$ labels the different copies of $\M_L$ in $(\M_L)^w$. When acting on a state in the $\mathbb{Z}_w$ twisted sector, the operators $V_\Delta^{(j)}(x)$ satisfy the property
\be
\label{monod}
V_\Delta^{(j)}(e^{2\pi i}x)=V_\Delta^{(j+1)}(x)~,
\ee
where $j=1,\ldots,w$, and $V_\Delta^{(w+1)}=V_\Delta^{(1)}$. Equation \eqref{monod} implies that the operator \eqref{untwistop} is single valued on the $x$-plane. It also means that the operators $V_\Delta^{(j)}(x)$ are invariant under $x\to e^{2\pi i w}x$. Therefore, it is natural to introduce the covering space, labeled by the coordinate~$t$ (see e.g. \cite{Roumpedakis:2018tdb} and references therein),
\be
\label{deft}
t=x^{1\over w}~.
\ee
When $x$ winds around the origin $w$ times, $t$ winds around the origin once. Thus, the $w$ copies of $\M_L$ that live on the $x$-plane give rise on the $t$-plane to a single copy of $\M_L$. The untwisted sector operator \eqref{untwistop} corresponds on the $t$-plane to the operator $V_\Delta$ in $\M_L$ that went into its construction. 

States in $(\M_L)^w/\mathbb{Z}_w$ inserted at $x=0$ (as in \eqref{stateV}) correspond on the $t$-plane to states in $\M_L$ inserted at $t=0$. In particular, a state $|V_{\Delta}\rangle$ on the $t$-plane corresponds on the $x$-plane to a state in the $\mathbb{Z}_w$ twisted sector, $|{\cal V} _\Delta^w\rangle$,  whose dimension is
\be 
\label{twistw}
\Delta_w = {\Delta - 1 \over w} + w~. \ee
An interesting subset of the correlation functions in $(\M_L)^w/\mathbb{Z}_w$ is 
\be 
\label{cortwist}
\langle{\cal V}_{\Delta_{l+1}}^w| \V_{\Delta_l}(x_l) \ldots \V_{\Delta_{1}}(x_{1}) |{\cal V}_{\Delta_0}^w\rangle~,\ee
where the operators $\V_{\Delta_j}(x)$ are the untwisted sector operators \eqref{untwistop}. The correlation functions \eqref{cortwist} describe the dynamics in a sector with given $w$. For $w=1$, they reduce to the correlation functions studied in the previous subsection. 

In that analysis, we found it convenient to express the results in terms of the modes of $V_\Delta(x)$, $V_{\Delta,n}$ \eqref{modeexpcont}, see \eqref{ointmodeV}. It is natural to generalize that construction to $w>1$. Thus, we define 
\be
\label{modew}
\V_{\Delta,n}=\oint{dx\over 2\pi i} x^{n+\Delta-1}\V_\Delta(x)~,
\ee
and compute the analog of \eqref{ointmodeV} for this case. In order to perform this calculation, we want to relate it to one in the CFT $\M_L$ in the covering space. To do that, we use the relation between $\V_\Delta(x)$ and $V_\Delta(t)$ discussed above,
\be
\label{nuv}
\V_{\Delta}(x)dx^\Delta=V_\Delta(t)dt^\Delta \ . 
\ee
Plugging \eqref{deft} and \eqref{nuv} into \eqref{modew}, we find that 
\be
\label{modenuv}
\V_{\Delta,n}=w^{1-\Delta}V_{\Delta,wn} \ . 
\ee
Thus, the momentum space analog of \eqref{cortwist} in the $\mathbb{Z}_w$ twisted sector in $(\M_L)^w/\mathbb{Z}_w$ can be expressed in terms of the following correlator in $\M_L$: 
\be\label{reltx} 
\langle{\cal V}_{\Delta_{l+1}}^w| \V_{\Delta_l,n_l} \cdots \V_{\Delta_1,n_1} |{\cal V}_{\Delta_0}^w\rangle =
\langle V_{\Delta_{l+1}}| V_{\Delta_l,wn_l} \cdots V_{\Delta_1,wn_1} | V_{\Delta_0}\rangle  \ . 
\ee
In \eqref{reltx}, we omitted an overall numerical constant that does not depend on the energies $n_j$. In the next section, we will reproduce this equation from the dual string theory perspective.

\section{Correlation functions in the $(0,2)$ string \label{sec3}}

The goal of this section is to reproduce the results of section \ref{sec2} in the $\N=2$ heterotic string. We will use the map between the string theory and CFT observables proposed in \cite{Giveon:2024fhz}. In particular, the state \eqref{vstate} corresponds in the $\N=2$ string to a state described by the vertex operator (see \cite{Giveon:2024fhz} for the notation)
\begin{equation}
	\label{fullver}
	O_\Delta=\int d^2z e^{-\bar\varphi_+-\bar\varphi_-}e^{-iEt}e^{ip_Lx_L+ip_Rx_R}V_\Delta~.
	\end{equation}
The physical state conditions (3.1), (3.4), (3.5) in \cite{Giveon:2024fhz} set 
\begin{equation}
	\label{massplus}
	E=p_R=\frac{n}{R}+\frac{wR}{\alpha'}~,\,\,\;\;\;\;p_L=\frac{n}{R}-\frac{wR}{\alpha'}~,\;\;\;\;\Delta-1=nw~.
	\end{equation}
The untwisted sector of the orbifold of section \ref{sec2}, described in subsection \ref{subsec21}, corresponds to $w=1$. We will discuss that case first, and then move on to $w>1$, which corresponds to the $\mathbb{Z}_w$ twisted sector of the orbifold, that we discussed in  subsection \ref{subsec22}. 

As a check, for $w=1$ the state \eqref{fullver} has spacetime scaling dimension $L_0=n+1=\Delta$ (see equation (3.37) in \cite{Giveon:2024fhz}), in agreement with the dimension of the state \eqref{vstate}. Note that the dimension $\Delta$ in \eqref{vstate} is a {\it spacetime} scaling dimension, while the operator $V_\Delta$ in \eqref{fullver} is a {\it worldsheet} operator of scaling dimension $\Delta$. As discussed in \cite{Giveon:2024fhz}, the worldsheet and spacetime properties are closely related in the $(0,2)$ string. This relation will play an important role below.
Note also that the $n$ in equation \eqref{massplus} is not directly related to the $n$ in equations \eqref{modeexp} and \eqref{modeexpcont}.

The two-point function \eqref{twoptf} and \eqref{normV} in the spacetime CFT corresponds in the $(0,2)$ string to the two-point function $\langle O^*_\Delta O_\Delta\rangle$. Recall that in $N>2$ point functions in string theory, we drop the $z$ integrals from three of the vertex operators, to account for the volume of the Conformal Killing Group (CKG) of the sphere, and integrate over the locations of the remaining $N-3$ operators. 

For $N=2$, dropping the integrals over the two vertex operators does not completely fix the CKG. Indeed, if we place the two operators at $z=0$ and $\infty$ on the worldsheet, there is still a subgroup of the CKG that remains unfixed, $z\to az$. The volume of this subgroup is infinite. A useful way of thinking about it is to map the sphere to a cylinder, with the two operators $O_\Delta$, $O^*_\Delta$ placed in the far past and far future, specifying the initial and final states. The infinite volume of the unfixed part of the CKG is proportional to the length of the worldsheet cylinder connecting the two.  

Thus, superficially, the two-point function $\langle O^*_\Delta O_\Delta\rangle$ vanishes, since we need to divide the finite worldsheet correlation function by the infinite volume of the residual CKG. However, as pointed out in \cite{Erbin:2019uiz}, the precise statement is that what vanishes is the term in the two-point function proportional to the length of the spacetime cylinder. There is a finite term proportional to the ratio of the lengths of the spacetime and worldsheet cylinders. This ratio takes the form $\infty/\infty$, and it was shown in \cite{Erbin:2019uiz} to give an unambiguous finite answer, which is proportional to the energy of the state created by $O_\Delta$. 

Using the results of \cite{Erbin:2019uiz}, we thus find that 
\be\label{twoptst}
\langle O^*_\Delta O_\Delta\rangle\simeq DE~,
\ee
where we used the worldsheet analog of \eqref{twoptf}, $E$ is the energy \eqref{massplus}, with $w=1$, and we omitted a numerical constant that can be read off from the analysis of \cite{Erbin:2019uiz}. Equation \eqref{twoptst} is the worldsheet analog of \eqref{normV}. It can be used to compare the normalization of the worldsheet state corresponding to the vertex operator $O_\Delta$ and the spacetime state \eqref{vstate}. 

Our next task is to calculate the $l+2$ point function \eqref{ointmode} from the worldsheet point of view. As explained in \cite{Giveon:2024fhz}, around equation (3.16), given a worldsheet $U(1)$ current $J(z)$, we can construct (the modes of) a spacetime current $K(x)$ as follows:
\begin{equation}
	\label{contour}
	K_n=\oint {dz\over 2\pi i} J(z) e^{i{n\over R}x^-(z,\bar z)}~.
	\end{equation}
Thus, the worldsheet correlator that corresponds to \eqref{ointmode} is given by 
\be
\label{wsointmode}
\oint_{\gamma_1}{dz_1\over2\pi i}\cdots \oint_{\gamma_l}{dz_l\over2\pi i}
\langle O^*_\Delta(\infty)
J(z_1) e^{i{n_1\over R}x^-(z_1)}\cdots 
J(z_l) e^{i{n_l\over R}x^-(z_l)}
O_\Delta(0)\rangle~.
\ee
In \eqref{wsointmode}, we dropped the integrals over the positions of the operators $O_\Delta$, $O^*_\Delta$ in \eqref{fullver}, and sent their positions on the worldsheet to zero and infinity, respectively. We also used the fact that, as discussed in \cite{Giveon:2024fhz}, in the string theory, {\it states} in $\M_L$ are described by vertex operators with $w=1$, such as \eqref{fullver}, while {\it operators} in this theory are described by $w=0$ vertex operators, such as \eqref{contour}. The contours $\gamma_j$ in \eqref{wsointmode} will be discussed shortly. 

The expectation value in \eqref{wsointmode} is taken in the worldsheet theory. The contribution of the worldsheet fields $x^\pm=t\pm x$ (defined in equation (2.9) in \cite{Giveon:2024fhz}) can be readily evaluated, and gives (for $w=1$)
\be \prod_{i=1}^l z_i^{n_i} \ .  \label{xsector} \ee
Plugging this into \eqref{wsointmode} leads to 
\be
\label{wsmatter}
\oint_{\gamma_1}{dz_1\over2\pi i}\cdots \oint_{\gamma_l}{dz_l\over2\pi i}
z_1^{n_1}\cdots z_l^{n_l}
\langle V_\Delta|
J(z_1) \cdots 
J(z_l) |V_\Delta\rangle~.
\ee
Comparing \eqref{wsmatter} to \eqref{ointmode}, we see that the two are closely related. Indeed, the subscript $n$ in \eqref{contour} is the same as that in \eqref{modeK}. Thus, the indices $n_i$ in \eqref{ointmode} are the same as $n_i$ in \eqref{wsmatter}. The two calculations are related by the map from the worldsheet to the spacetime $x(z)=z$. Hence, the contours $\gamma_i$ in the $z$-plane are the same as the ones we took in \eqref{ointmode}. 

To actually calculate \eqref{wsmatter}, we write it in terms of the modes of the {\it worldsheet} current algebra (the worldsheet analog of \eqref{matlmode}), and commute all the annihilation operators $J_{n>0}$ through the creation operators $J_{n<0}$, using the fact that they annihilate the state created by $O_\Delta(0)$. The result of this calculation is identical to the one obtained in spacetime using \eqref{matlmode}, since the modes $J_n$ of $J(z)$, the current defined around equation \eqref{contour}, satisfy the same commutation relations as the spacetime modes $K_n$, defined in \eqref{modeK}, \cite{Giveon:2024fhz}. Remembering the rescalng factor between the worldsheet and spacetime two-point functions \eqref{twoptf} and  \eqref{twoptst}, we find that all the $l+2$ point functions agree.

For general positive $w$, (\ref{xsector}) takes the form
\be \prod_{i=1}^l z_i^{wn_i} \label{xsector2}~. \ee
Thus, the amplitude \eqref{wsointmode} takes in this case the form 
\be
\label{genw}
\langle O^*_\Delta(\infty)
J_{wn_1} \cdots J_{wn_l}
O_\Delta(0)\rangle~.
\ee
According to the map proposed in \cite{Giveon:2024fhz}, it should be compared to the right hand side of \eqref{reltx} (for the particular case where all the operators $V_{\Delta_j}$ are taken to be the currents $K$ given in \eqref{contour}). Thus, we learn that the string theory calculation gives the same answer as the dual CFT one for all~$w$. 

An interesting feature of the analysis of this section is that the worldsheet coordinate $z$ plays the same role in the string theory calculation as the covering space coordinate $t$ in section \ref{subsec22} played in the dual orbifold CFT one. This is one of many features that our system shares with string theory on $AdS_3$ (for further discussion, see \cite{Giveon:2024fhz}).

Note also that we restricted in the above discussion to amplitudes in which the $w=0$ operators are KM currents, \eqref{contour} and \eqref{wsointmode}, which correspond in the dual theory to \eqref{ointmode}. It is not hard to generalize to the case where they are operators corresponding to general Virasoro primaries in the dual CFT, $V_\Delta$. The corresponding vertex operators were found in \cite{Giveon:2024fhz}, and are given by equation (3.45) in that paper. Inserting them into the analog of \eqref{wsointmode} gives the worldsheet version of the correlation functions \eqref{ointmodeV} and \eqref{reltx}.  

As an example of this procedure, consider the string theory analog of the calculation \eqref{excorl} -- \eqref{modeexm}. According to the rules described in \cite{Giveon:2024fhz}, the corresponding string theory amplitude is  
\be
\label{stpamp}
\langle O^*(\vec p_3) \V_n(\vec p_2)O(\vec p_1)\rangle~,
\ee
where 
\be
\label{defop}
O(\vec p)=\int d^2z e^{-\bar\varphi_+-\bar\varphi_-}e^{-iEt}e^{ip_Lx_L+ip_Rx_R}e^{i\vec p\cdot \vec y}~,
\ee
and $E$, $p_L$, $p_R$ are given by \eqref{massplus}, with $w=1$ and $\Delta=\Delta_{\vec p}=\frac12|\vec p|^2$. The $w=0$ vertex operator $\V_n(\vec p)$ is given by equation (3.45) in \cite{Giveon:2024fhz},
\begin{equation}
	\label{higherS}
	\V_n(\vec p)=\oint_\gamma \frac{dz}{2\pi i} 
 {e^{i\vec p\cdot\vec y}\over (\partial x^-)^{\Delta_{\vec p}-1}}
 e^{i\frac nR x^-}\ .
	\end{equation}
We next verify that plugging \eqref{defop} and \eqref{higherS} into \eqref{stpamp} gives \eqref{modeexm}. 

As discussed above, we start by stripping off the integrals over the worldsheet positions of the two $(-1,-1)$ picture operators, to fix part of the CKG of the sphere. This also allows us to place the vertex operators $O(\vec p_1)$ and $O^*(\vec p_3)$ at $z=0$ and $\infty$, respectively. The remaining part of the CKG is treated as in the discussion leading to equation \eqref{twoptst}. 

The resulting three-point function can be written as a product of three factors. The ghost contribution is trivial, and we will not write it. The remaining two contributions come from the worldsheet Niemeier CFT, and from the spacetime fields $x^\pm$. The former takes the form 
\be
\label{niemcont}
\langle e^{-i\vec p_3\cdot \vec y(\infty)}
e^{i\vec p_2\cdot \vec y(z)}e^{i\vec p_1\cdot \vec y(0)}\rangle=\frac{\delta_{\vec p_1+\vec p_2,\vec p_3}}{z^{\Delta_{12}}}~,
\ee
where $\Delta_{12}=\Delta_1+\Delta_2-\Delta_3=-\vec p_1\cdot\vec p_2$. The latter is given by 
\be
\label{xpmcont}
\langle 
e^{iE_3t}e^{-ip_{3,L}x_L-ip_{3,R}x_R}(\infty)
{e^{i\frac nR x^-}\over (\partial x^-)^{\Delta_2-1}}(z)
e^{-iE_1t}e^{ip_{1,L}x_L+ip_{1,R}x_R}(0)
\rangle=z^{\Delta_{12}-1}\delta_{n,\Delta_1-\Delta_3}~,
\ee
where we again omitted some overall numerical constants. Note that the $z$ dependence on the right hand side of \eqref{xpmcont} is compatible with the fact that the three operators have worldsheet dimensions $1-\Delta_i$. The value of $n$ is determined by momentum conservation, which gives the Kronecker $\delta$ in \eqref{xpmcont}. 

Plugging \eqref{niemcont} and  \eqref{xpmcont} into \eqref{stpamp}, and performing the contour integral \eqref{higherS} over a contour surrounding the origin, gives the result \eqref{modeexm}. The numerical constants we did not keep track of can be used to fix the relative normalizations of the states and operators in the string theory and the dual CFT.  

The above calculation can be generalized to $w>1$, and shown to agree with \eqref{reltx} with $l=1$. In particular, the factor $w^{1-\Delta}$ in \eqref{modenuv} follows from the power of $\partial x^-$ in \eqref{higherS}. The calculation can also be generalized to $l>1$ in \eqref{ointmodeV}.

\section{Discussion}

In sections \ref{sec2} and \ref{sec3}, we established a map between a certain class of correlation functions in the symmetric orbifold CFT, $\left(\M_L\right)^N/S_N$, and its string theory dual. In the CFT, this class involves correlation functions of untwisted sector operators in a state that belongs to the $\mathbb{Z}_w$ twisted sector, while in the string theory it involves correlation functions of short string operators in a state that corresponds to a single string with winding $w$ around the spatial circle.  

It is natural to ask how this map extends to more general amplitudes. In the string theory language, these are correlators with more than two operators with non-zero winding. A class of such correlators that is easy to compute can be constructed as follows. Looking back at the vertex operators \eqref{fullver}, we note that for $\Delta=1$ and $n=0$, there are such operators for any $w>0$. One can think of them as creating the ground states in the sectors with given $w$ (see \eqref{twistw}). The worldsheet operators $V_\Delta(z)$ correspond in this case to worldsheet currents $J^a(z)$, $a=1,\cdots, {\rm dim}\; G$, as discussed in detail in \cite{Giveon:2024fhz}. In the $(0,0)$ picture, the corresponding spacetime operators take the form $\widetilde K^a_{-w}$, with
\begin{equation}
	\label{tildeK}
	\widetilde K^a_w=\oint {dz\over 2\pi i} J^a(z) e^{i{wR\over \alpha'}\left(x_L^+(z)+x_R^-(\bar z)\right)}~.
	\end{equation}
This should be compared to \eqref{contour}, which involves $x^-(z,\bar z)=x_L^-(z)+x_R^-(\bar z)$. Clearly, the operators $K_n$, \eqref{contour}, and $\widetilde K_w$ are related by T-duality on the spatial $(x)$ circle. Thus, the correlation functions \eqref{wsointmode} are related by T-duality to those containing two arbitrary operators \eqref{fullver} with any number of operators of the form  \eqref{tildeK}. 

The operators \eqref{tildeK} play an interesting role in the spacetime CFT. Repeating the analysis of the operators $K^a_n$ in \cite{Giveon:2024fhz}, it is easy to show that they also form a $\widehat G$ affine Lie algebra, 
\begin{equation}
	\label{commutrel}
	[\widetilde K^a_w, \widetilde K^b_{w'}]=if^{ab}_{\,\,\,\,\,\,c}\widetilde K^c_{w+w'}+
 \frac 12 w\delta^{ab}\delta_{w+w',0}P^+_L~,
	\end{equation}
where 
\begin{equation}
	\label{plm}
	P_L^+=\frac{R}{\alpha'}\oint\frac{dz}{2\pi i} i\partial x^+ \ .
	\end{equation}
However, while the $K^a_n$ act on a sector with given winding $w$, the $\widetilde K^a_w$ change the winding by the amount $w$, and instead act in a sector with given momentum $n$. While the level of the affine Lie algebra of the $K$'s is $w$, the one of the $\widetilde K$'s is $n$. The ground states in a sector with given $w>1$ can be thought of as the states 
$$\widetilde K^a_{-w}|0\rangle~,$$
where $|0\rangle$ is the $SL(2,\mathbb{R})$ invariant vacuum of the full spacetime theory. 

In \cite{Giveon:2024fhz}, we showed that the operators $K^a_{-n}$ can be thought of as spectrum generating operators in the seed CFT $\M_L$. The operators $\widetilde K^a_{-w}$ play a similar role in the spacetime CFT \eqref{mln}. However, unlike the $K$'s, they connect sectors with different windings. Thus, it is natural to expect that by acting with the creation operators of both $K^a$ and $\widetilde K^a$ on primaries of both algebras, we can generate the full spectrum of the symmetric product theory \eqref{mln}. We will leave the analysis of the resulting algebraic structure to future work, restricting here to some brief comments.  

A first step towards analyzing the algebraic structure of the theory is to calculate the commutation of the $K$'s, \eqref{contour}, and $\widetilde K$'s, \eqref{tildeK}. A straightforward calculation shows that for $nw<-1$ the commutator $[ K^a_n,\widetilde K^b_w]$ vanishes, but when $n,w$ are both positive or negative it is non-zero, and gives rise to additional charges that are labeled by $(n,w)$. We will leave a full analysis of the resulting symmetry algebra to future work. 

The above analysis was done in the $(0,2)$ string theory, however, we expect it to apply to the CFT \eqref{mln}. In particular, that CFT should contain the Kac-Moody generators \eqref{tildeK}. It would be interesting to construct them directly, presumably by applying general studies of twisted sector operators, such as \cite{Lunin:2000yv,Dei:2019iym}, to the symmetric product of Niemeier CFT's.  

We finish this section with some comments about our construction. 
\begin{itemize}
\item Here and in \cite{Giveon:2024fhz}, we were led to study two types of physical vertex operators. One is vertex operators, such as \eqref{fullver}, that create states with non-zero winding when acting on the vacuum. The other, such as \eqref{higherS}, correspond to operators with zero winding, that can be thought of as probing these states. The resulting picture is reminiscent of D-brane physics (see e.g. \cite{Hashimoto:1996bf,Witten:1998qj,Gubser:1998bc,Gubser:1998kv,Hashimoto:2001sm} and references therein for relevant papers in that context). The wound strings created by \eqref{fullver} are analogs of the boundary states describing D-branes, while \eqref{higherS} are analogous to closed string operators that probe these states. 
\item The two types of operators mentioned above are also analogous to what happens in string theory on $AdS_3$. For example, in the construction of \cite{Balthazar:2021xeh} of vacua with $R_{AdS}<l_s$, the dual CFT is a (deformed) symmetric product. The states in the symmetric product correspond to vertex operators with $w>0$, and the sector with $w=0$ contains the symmetry generators of the spacetime CFT, like in the $\N=2$ heterotic string. 
\item It is important to emphasize that, as mentioned above, the string theory correlation functions that we computed, e.g.\ 
\eqref{twoptst} and \eqref{wsmatter}, are not the standard $S$-matrix elements one usually computes in string theory. The latter compute the term that is proportional to the length of time in the correlation function, which is interpreted as the statement that the scattering process can occur at any time, in a translationally invariant theory. In the correlation functions computed in this paper, that term vanishes but, as described above, there is a finite term that maps in the CFT \eqref{mln} to a correlation function of the form \eqref{ointmode} and \eqref{ointmodeV}. 
\end{itemize}

\section*{Acknowledgements}

The work of AG was supported in part by the ISF (grant number 256/22).
The work of AH was supported in part by the U.S. Department of Energy, Office of Science, Office of High Energy Physics, under Award Number DE-SC0017647.
The work of DK was supported in part by DOE grant DE-SC0009924.

\providecommand{\href}[2]{#2}\begingroup\raggedright\endgroup


\begin{thebibliography}{10}

\bibitem{Giveon:2024fhz}
A.~Giveon, A.~Hashimoto, and D.~Kutasov, ``{${\cal N}=2$ Heterotic Strings
  Revisited},'' \href{http://www.arXiv.org/abs/2409.18183}{{\tt 2409.18183}}.

\bibitem{Aharony:2024fid}
O.~Aharony and E.~Y. Urbach, ``{Type II string theory on
  AdS3\texttimes{}S3\texttimes{}T4 and symmetric orbifolds},'' {\em Phys. Rev.
  D} {\bf 110} (2024), no.~4, 046028,
  \href{http://www.arXiv.org/abs/2406.14605}{{\tt 2406.14605}}.

\bibitem{Marcus:1992wi}
N.~Marcus, ``{A Tour through N=2 strings},'' in {\em {International Workshop on
  String Theory, Quantum Gravity and the Unification of Fundamental
  Interactions}}.
\newblock 11, 1992.
\newblock \href{http://www.arXiv.org/abs/hep-th/9211059}{{\tt hep-th/9211059}}.

\bibitem{Roumpedakis:2018tdb}
K.~Roumpedakis, ``{Comments on the S$_{N}$ orbifold CFT in the large
  $N$-limit},'' {\em JHEP} {\bf 07} (2018) 038,
  \href{http://www.arXiv.org/abs/1804.03207}{{\tt 1804.03207}}.

\bibitem{Erbin:2019uiz}
H.~Erbin, J.~Maldacena, and D.~Skliros, ``{Two-Point String Amplitudes},'' {\em
  JHEP} {\bf 07} (2019) 139, \href{http://www.arXiv.org/abs/1906.06051}{{\tt
  1906.06051}}.

\bibitem{Lunin:2000yv}
O.~Lunin and S.~D. Mathur, ``{Correlation functions for M**N / S(N)
  orbifolds},'' {\em Commun. Math. Phys.} {\bf 219} (2001) 399--442,
  \href{http://www.arXiv.org/abs/hep-th/0006196}{{\tt hep-th/0006196}}.

\bibitem{Dei:2019iym}
A.~Dei and L.~Eberhardt, ``{Correlators of the symmetric product orbifold},''
  {\em JHEP} {\bf 01} (2020) 108,
  \href{http://www.arXiv.org/abs/1911.08485}{{\tt 1911.08485}}.

\bibitem{Hashimoto:1996bf}
A.~Hashimoto and I.~R. Klebanov, ``{Scattering of strings from D-branes},''
  {\em Nucl. Phys. B Proc. Suppl.} {\bf 55} (1997) 118--133,
  \href{http://www.arXiv.org/abs/hep-th/9611214}{{\tt hep-th/9611214}}.

\bibitem{Witten:1998qj}
E.~Witten, ``{Anti-de Sitter space and holography},'' {\em Adv. Theor. Math.
  Phys.} {\bf 2} (1998) 253--291,
  \href{http://www.arXiv.org/abs/hep-th/9802150}{{\tt hep-th/9802150}}.

\bibitem{Gubser:1998bc}
S.~S. Gubser, I.~R. Klebanov, and A.~M. Polyakov, ``{Gauge theory correlators
  from noncritical string theory},'' {\em Phys. Lett. B} {\bf 428} (1998)
  105--114, \href{http://www.arXiv.org/abs/hep-th/9802109}{{\tt
  hep-th/9802109}}.

\bibitem{Gubser:1998kv}
S.~S. Gubser, A.~Hashimoto, I.~R. Klebanov, and M.~Krasnitz, ``{Scalar
  absorption and the breaking of the world volume conformal invariance},'' {\em
  Nucl. Phys. B} {\bf 526} (1998) 393--414,
  \href{http://www.arXiv.org/abs/hep-th/9803023}{{\tt hep-th/9803023}}.

\bibitem{Hashimoto:2001sm}
A.~Hashimoto and N.~Itzhaki, ``{Observables of string field theory},'' {\em
  JHEP} {\bf 01} (2002) 028,
  \href{http://www.arXiv.org/abs/hep-th/0111092}{{\tt hep-th/0111092}}.

\bibitem{Balthazar:2021xeh}
B.~Balthazar, A.~Giveon, D.~Kutasov, and E.~J. Martinec, ``{Asymptotically free
  AdS$_{3}$/CFT$_{2}$},'' {\em JHEP} {\bf 01} (2022) 008,
  \href{http://www.arXiv.org/abs/2109.00065}{{\tt 2109.00065}}.

\end{thebibliography}
\end{document}